\documentclass[intlimits,twoside,a4paper]{article}

\usepackage{amsmath, amssymb}
\usepackage{wrapfig}
\usepackage[dvips]{graphicx}

\usepackage[T2A]{fontenc}
\usepackage[cp1251]{inputenc}

\usepackage{cmpj2}

\issue{2013}{16}{3}{33702}
\doinumber{10.5488/CMP.16.33702}

\title[Effect of quantum dot shape]%
{Effect of quantum dot shape of the GaAs/AlAs heterostructure on interlevel hole light absorption}%

\author[V.I.~Boichuk \textsl{et al.}]{V.I.~Boichuk, I.V.~Bilynskyi, O.A.~Sokolnyk\thanks{E-mail: sokolnik@gmail.com}\ , I.O.~Shakleina}
\address{Ivan Franko Drohobych State Pedagogical University, Theoretical Physics Department, \\
3 Stryiska St., 82100 Drohobych, Ukraine}

\date{Received October 2, 2012, in final form February 2, 2013}
\authorcopyright{V.I.~Boichuk, I.V.~Bilynskyi, O.A.~Sokolnyk, I.O.~Shakleina, 2013}

\begin{document}

\maketitle

\begin{abstract}
The effect of quantum dot shape on the hole energy spectrum and optical properties caused by the interlevel charge transition based on the $4\times4$ Hamiltonian has been studied for the GaAs quantum dot in the AlAs semiconductor matrix. Calculations have been carried out in perturbation theory taking into account the hybridization of states for cubic, ellipsoidal, cylindrical and tetrahedral shapes by changing the volume of a quantum dot. Based on the energy calculations and on the determined wave functions of the hole states we have defined the selection rules and the dependence of the interlevel hole absorption coefficient on QD shapes and volumes.

\keywords multiband approximation, optical transitions, oscillator strength, absorption coefficient

\pacs 71.15.-m, 78.20.Ci, 78.67.Hc
\end{abstract}

\section*{Introduction}

Lately, quantum dots (QDs) have attracted the attention of researchers both as an object of the study of nanoheterostructure physical properties and as an opportunity for different applications of QDs in everyday life.

As concerns the practical use, QDs are used not only in solar batteries, where the shape and size of CdSe/TiO$_2$ QDs make it possible to improve the photoresponse~\cite{Kong}, but also for the comparison with organic dyes as fluorescent labels based on luminescence and fluorescence~\cite{Resch}.

Nowadays, various fields of science use the achievements of physics of nanosize heterostructures, quantum dots in particular. QDs are actively used in the advanced medical technology and cytology~\cite{Tan}. The effect of QD shape is of great importance in these processes. The recent researches show that the shape of quantum dots affects CdTe quantum dot phagocytosis~\cite{Lu1}.

As noted above, the shape has a significant effect on the QD physical properties. For example, the authors of~\cite{Krasnok} proved that the shape of a quantum dot affects not only the energy of charges in QDs, but also their electronic density of states. The effect of the shape on the piezoelectric effect and electronic properties of In(Ga)As/GaAs quantum dots, as well as on the longitudinal photoconductivity of multilayer Ge/Si structures with Ge quantum dots has also been visibly demonstrated~\cite{Schli, Taloch}.

Recent experimental investigations reveal the effect of QD shape on the structure and arrangement of Si/Ge quantum dots in the arrays that are used in nanoelectrical devices~\cite{Dais}. The role of QD shape is determinant in obtaining various shapes and sizes of a QD by precipitation~\cite{Lu2}.

The goal of modern theoretical research is to bring the results to the best agreement with experimental data. Therefore, while calculating physical parameters of semiconductor nanoheterostructures, one should take into account a real band crystal structure of the nanosystem~\cite{Moskal, Linnik}.

Applying the envelope-function approximation to the description of hole states in QDs embedded in a semiconductor matrix, we can find not only the energy spectrum and wave functions of particles, but also examine the optical effects caused by interlevel hole transitions~\cite{Li, Kupc}. It is known that the optical selection rules, oscillator strength, and interlevel hole absorption coefficient essentially depend on the QD size and shape~\cite{Polup, Menen, Fonob}.

The aim of the present work is to study the effect of QD shape on the hole energy spectrum and on the optical parameters (the hole transition probability, transition oscillator strength, interband light absorption coefficient) of a semiconductor nanoheterosystem. QDs of wide energy-band gap semiconductors have been considered and exact solutions of the Schr\"odinger equation for a hole in a spherical QD in the four-band effective-mass approximation have been used. QDs of other shapes (cubic, ellipsoidal, cylindrical, tetrahedral) have been chosen to be of the same volume and in such a way that the potential energy of carriers for the QD shape of interest should be close to its potential energy in a spherical QD. The calculations have been carried out for the GaAs QD which is placed into the AlAs matrix.

\section{The energy and wave functions of bound states of holes in a QD}

We study a semiconductor QD embedded in a semiconductor or dielectric matrix with a larger energy gap. If the QD has a spherical shape, for many semiconductor systems the problem of finding the hole energy spectrum (under certain assumptions) is solved analytically.

We consider such crystals of a heterostructure in which the Brillouin zone center is a point of the valence band degeneracy. This valence band structure is observed in most semiconductors. Let us consider the semiconductor heterostructures in which the spin-orbit interaction does not significantly affect the basic hole properties. This can be done if the parameter of this interaction is sufficiently large. Then, the spin-orbit band can be neglected.

Ignoring the warping of the isoenergetic surfaces in the $k$-space (spherical approximation), the hole Hamiltonian can be presented as~\cite{Balder}
\begin{equation} \label{EQ__1_}
\hat{H}_{0} =\frac{1}{2} \left(\gamma _{1} +\frac{5}{2} \gamma \right)\hat{p}^{2} -\gamma \left(\hat{\vec{p}}\cdot \hat{\vec{J}}\right)^{2} +U_\textrm{s} \left(r\right), \end{equation}
\noindent where $\gamma ,\gamma _{1} ,\, \, \gamma _{2} ,\, \, \gamma _{3} $ are the Luttinger parameters [where $\gamma =1/5\left(3\gamma _{3} +2\gamma _{2} \right)$], $\hat{\vec{J}}$ is the spin momentum operator, for which $J^{2} =\hbar ^{2} j(j+1),$  $j=3/2 $, $\hat{\vec{p}}$ is the quasiparticle momentum operator, $U_\textrm{s} \left(r\right)$ is the charge potential energy in a spherical QD.

The wave function, which is a solution of the equation with the Hamiltonian~\eqref{EQ__1_}, is expressed as a product of the total momentum operator eigenfunctions and radial functions~\cite{Gelm}. As shown in~\cite{Balder, Lipari}, it is convenient to divide the solutions obtained with this wave function into even and odd states, (I, II). Taking into account the states with $f=3/2, 5/2 $, we choose them as follows:
\begin{equation}
	\label{EQ__2_}
	v _{f,L,M}\left(r,\theta ,\phi \right) = \left\{
	\begin{array}{ll}
	{v _{f,L,M}^\textrm{I}  \left(r,\theta ,\phi \right) = R_{2}^\textrm{I}  \left(r\right) \Phi _{f,M}^{f-\frac{3}{2} }  \left(\theta ,\phi \right)+R_{1}^\textrm{I}  \left(r\right)\Phi _{f,M}^{f+\frac{1}{2} }  \left(\theta ,\phi \right), } & {\left(L=f-\frac{3}{2} , f+\frac{1}{2} \right),} \\[1ex]
	{v _{f,L,M}^\textrm{II}  \left(r,\theta ,\phi \right) = R_{2}^\textrm{II}  \left(r\right) {\Phi _{f,M}^{f-\frac{1}{2} }  \left(\theta ,\phi \right)}+R_{1}^\textrm{II}  \left(r\right)\Phi _{f,M}^{f+\frac{3}{2} }  \left(\theta ,\phi \right), } & { \left(L=f-\frac{1}{2} , f+\frac{3}{2} \right),}
	\end{array}
	\right.
\end{equation}
\noindent where $\Phi _{f,M}^{L} \left(\theta ,\phi \right)$ are the four-component spinors corresponding to spin $j=3/2 $. A general form of spinors can be also represented  in terms of the Clebsch-Jordan coefficients~\cite{Davyd}
\begin{equation} \label{EQ__3_} \Phi _{f,M}^{L} \left(\theta ,\phi \right)=\sum _{m=-L}^{L}\sum _{m_{j} =-j}^{j}G_{L,m;j,m_{j} }^{f,M} Y_{L,m} \left(\theta ,\phi \right)\xi _{m_{j} } \, . \end{equation}
\noindent where $ Y_{L,m} $ are spherical harmonics, $ \xi _{m_{j}} $ is the spinor function.

In the inner region of the heterosystem ($r\leqslant a$), the solutions of the corresponding equations can be written using the Bessel functions of the first kind
\begin{eqnarray}
\label{EQ__4_}
	R_1^{i,(1)}\left( r \right) &=& \frac{{B_1^{(1)}}}{{\sqrt r }}{J_{L + \frac{5}{2}}}\left( {\frac{{{k^{(1)}}r}}{{\sqrt {1 - 	
	{{\left( {{\mu ^{(1)}}} \right)}^2}} }}} \right) + \frac{{B_2^{(1)}}}{{\sqrt r }}{J_{L + \frac{5}{2}}}\left(
	{\frac{{{k^{(1)}}r}}{{\sqrt {1 + {{\left( {{\mu ^{(1)}}} \right)}^2}} }}} \right),
	\nonumber \\
	R_{2}^{i,(1)} \left(r\right)&=&\frac{B_{1}^{(1)} \left(C_{1}^{(1)} -\mu ^{(1)} \right)}
	{\sqrt{r} \sqrt{\left(\mu ^{(1)} \right)^{2} -\left(C_{1}^{(1)} \right)^{2} } }
	{ J}_{L+\frac{1}{2} } \left(\frac{k^{(1)} r}{\sqrt{1-\left(\mu ^{(1)} \right)^{2} } } \right)
	\nonumber \\
	&&{ +\frac{B_{2}^{(1)} \left(C_{1}^{(1)} +\mu ^{(1)} \right)}{\sqrt{r} \sqrt{\left(\mu ^{(1)} 	
	\right)^{2} -\left(C_{1}^{(1)} \right)^{2} } } {J}_{L+\frac{1}{2} } \left(\frac{k^{(1)} r}{\sqrt{1+\left(\mu ^{(1)}
	\right)^{2} } } \right),}
\end{eqnarray}
where
\[
k^{(1)} =\sqrt{2E/\gamma _{1}^{(1)} } , \qquad \lambda ^{(1)} =\sqrt{2E/\left[\gamma _{1}^{(1)} \left(1+\mu ^{(1)} \right)\right]}, \qquad \mu ^{(1)} =\frac{2\gamma ^{(1)} }{\gamma _{1}^{(1)} }, \qquad i=\textrm{I,\,II}.
\]
In the matrix for $r>a$, the solutions of the equations can be represented by modified Bessel functions of the second kind
\begin{eqnarray}
\label{EQ__5_}
	R_{1}^{i,(2)} \left(r\right)&=&\frac{D_{1}^{(2)} }{\sqrt{r} } K_{L+\frac{5}{2} } \left(\frac{k^{(2)} r}{\sqrt{1-\left
	(\mu ^{(2)} \right)^{2} } } \right)+\frac{D_{2}^{(2)} }{\sqrt{r} } K_{L+\frac{5}{2} }
	\left(\frac{k^{(2)} r}{\sqrt{1+\left(\mu ^{(2)} \right)^{2} } } \right),
	\nonumber \\
	R_{2}^{i,(2)} \left(r\right)&=&\frac{D_{1}^{(2)} \left(-C_{1}^{(2)} +\mu ^{(2)} \right)}{\sqrt{r} \sqrt{\left(\mu ^{(2)}
	 \right)^{2} -\left(C_{1}^{(2)} \right)^{2} } } K_{L+\frac{1}{2} } \left(\frac{k^{(2)} r}{\sqrt{1-\left(\mu ^{(2)}
	 \right)^{2} } } \right)
	\nonumber \\
	&&+\frac{D_{2}^{(2)} \left(-C_{1}^{(2)} -\mu ^{(2)} \right)}{\sqrt{r} \sqrt{\left(\mu ^{(2)} \right)^{2} -
	\left(C_{1}^{(2)} \right)^{2} } } K_{L+\frac{1}{2} } \left(\frac{k^{(2)} r}{\sqrt{1+\left(\mu ^{(2)}
	\right)^{2} } } \right),
\end{eqnarray}
where
\[
k^{(2)} =\sqrt{2\left(U_{0} -E\right)/\gamma _{1}^{(2)} } , \qquad \lambda ^{(2)} =\sqrt{2\left(U_{0} -E\right)/\left[\gamma _{1} ^{(2)} \left(1+\mu ^{(2)} \right)\right]} , \qquad \mu ^{(2)} =\frac{2\gamma ^{(2)} }{\gamma _{1}^{(2)} }, \qquad i=\textrm{I,\,II}.
\]

In order to match the solutions, two conditions have been used: the continuity of the radial wave function and the normal component of the probability density flux through a QD spherical surface~\cite{Boich}. Under such conditions one can define the wave functions of the hole states $v _{fLMn}$ ($n=1,\, 2,\, 3,\, \dots$ is the ordinal number of solution of the dispersion equation) and the energies for different values of the radius (volume) of the spherical QD. We will consider further QDs of cubic, ellipsoidal, cylindrical and pyramidal shapes having the volume identical with that of a spherical QD. In this case the potential $U(\vec{r})$ for a hole of the QD of any shape is not significantly different from the potential $U_\textrm{s} (r)$ of the charge in a spherical QD as is shown by calculations. Then, the Schr\"odinger equation can be presented in the form
\begin{equation} \label{EQ__6_} \left[\frac{1}{2} \left(\gamma _{1} +\frac{5}{2} \gamma \right)\hat{p}^{2} -\gamma \left(\hat{\vec{p}}\cdot \hat{\vec{J}}\right)^{2} +U_\textrm{s} \left(r\right)+W(\vec{r})\right]\psi (\vec{r})=E\psi (\vec{r}), \end{equation}
\noindent where $W(\vec{r})=U(\vec{r})-U_\textrm{s} (r)$ is the perturbation potential. The potential $W(\vec{r})$ has the symmetry of the considered QD, which is lower than the symmetry of a spherical QD.

Let us investigate the genesis of the hole quantum states by changing the QD shape. To achieve this aim, we introduce the wave function $\psi $ as a linear combination of functions $v _{fLMn} $~\eqref{EQ__2_} and find the energy spectrum of holes in real QD:
\begin{equation}
	\label{EQ__7_}
	\psi =\sum _{fLMn}C_{fLMn} \, v _{fLMn} \,.
\end{equation}

Equation~\eqref{EQ__6_} can be represented by a set of linear algebraic equations
\begin{equation} \label{EQ__8_} \sum _{f'L'M'n'}\langle fLMn|\hat{H}|f'L'M'n'\rangle C_{f'L'M'n'} =E\; C_{fLMn} \,, \end{equation}
\noindent where $\langle fLMn|\hat{H}|f'L'M'n'\rangle =\varepsilon _{fLMn} \delta _{f,f'}^{} \delta _{L,L'} \delta _{M,M'} \delta _{n,n'} +W_{fLMn,f'L'M'n'} $, $\varepsilon _{fLMn}$ is the corresponding energy of a spherical QD.

The diagonalization of the $\langle fLMn|\hat{H}|f'L'M'n'\rangle $ matrix applying a unitary transformation is required to find the hole spectrum
\begin{eqnarray}
\label{EQ__9_}
	&&\sum _{\begin{smallmatrix}{l} {f'L'M'n'} \\ {f''L''M''n''} \end{smallmatrix}}
 \langle fLMn|U\hat{H}U^{-1} |
	f''L''M''n''\rangle U_{f''L''M''n'',f'L'M'n'} C_{f'L'M'n'} =
	\nonumber \\
	&&\qquad\qquad\qquad=E\, \sum _{f'L'M'n'}U_{fLMn,f'L'M'n'} C_{f'L'M'n'} \; .
\end{eqnarray}
We get
\begin{equation} \label{EQ__10_} \sum _{f'L'M'n'}\left(\hat{H}_{fLMn,f'L'M'n'} \delta _{f,f'} \delta _{L,L'} \delta _{M,M'} \delta _{n,n'} -E\right)\tilde{C}_{f'L'M'n'} =0\,, \end{equation}
\noindent where $\hat{H}=U\hat{H}U^{-1} $ and $\tilde{C}_{f'L'M'n'} =\; \sum _{f'L'M'n'}U_{fLMn,f'L'M'n'} C_{f'L'M'n'} $.

\begin{figure}[!b]
\centerline{
\includegraphics[width=0.65\textwidth]{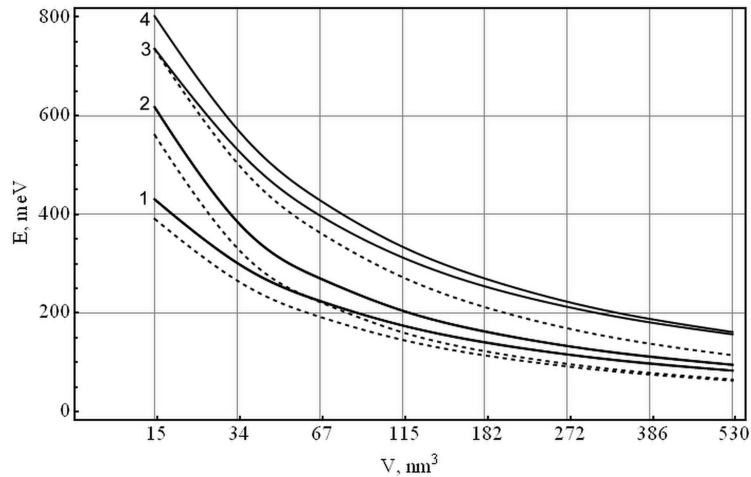}
}
	\caption{The hole energy of a cubic QD.}
	\label{fig1}
\end{figure}

\begin{figure}[!t]
\centerline{
\includegraphics[width=0.65\textwidth]{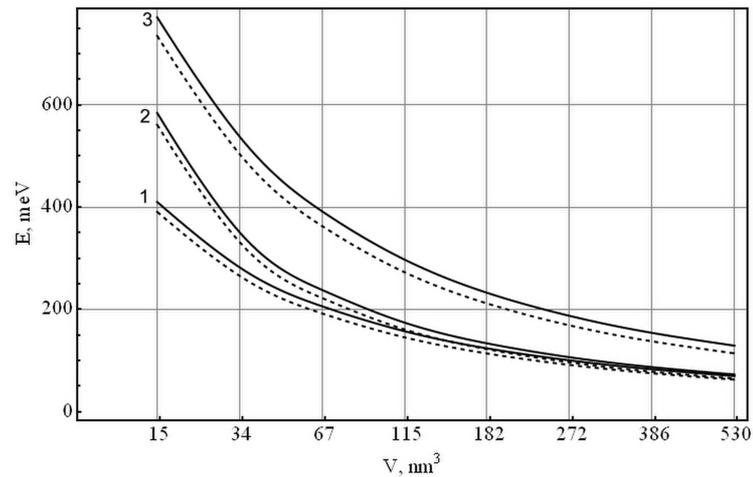}
}
	\caption{The hole energy of an ellipsoidal QD.}
	\label{fig2}
\end{figure}

\begin{figure}[!b]
\centerline{
\includegraphics[width=0.65\textwidth]{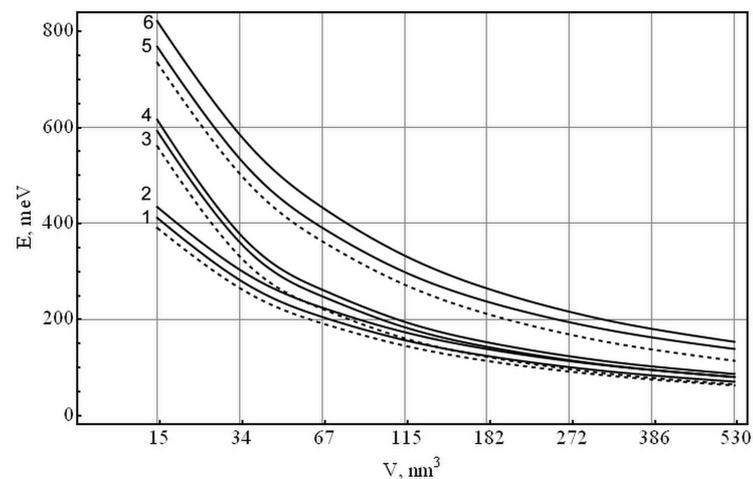}
}
	\caption{The hole energy of a cylindrical QD.}
	\label{fig3}
\end{figure}

\begin{figure}[!t]
\centerline{
\includegraphics[width=0.65\textwidth]{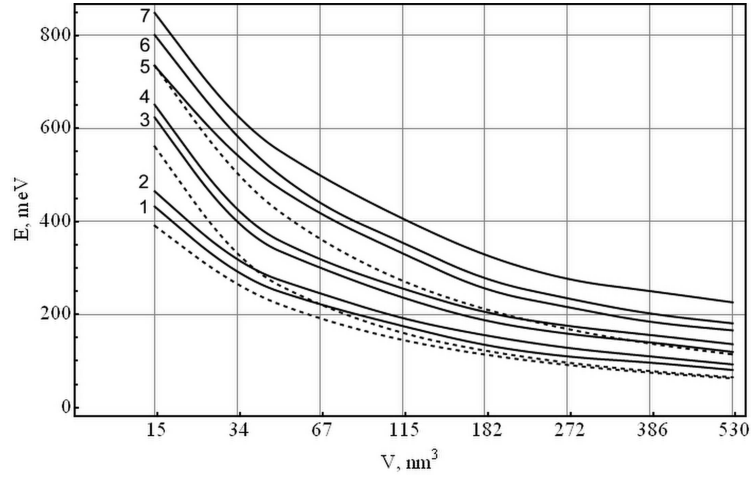}
}
	\caption{The hole energy of a pyramidal QD.}
	\label{fig4}
\end{figure}

Further on, we will consider in more detail the change of the hole energy for the three lowest levels in a spherical QD (dashed lines in figures~\ref{fig1}--\ref{fig4}) and in QDs of lower symmetry shape (solid lines). In the sum~\eqref{EQ__7_} we restrict ourselves to the wave functions of the first three levels of the spherical QD. Then, we have 14 different wave functions~\eqref{EQ__2_} and coefficients $\tilde{C}_{f'L'M'n'} $ , and the $\langle fLMn|\hat{H}|f'L'M'n'\rangle $ matrix contains 196 elements.

The calculations have been performed for the GaAs/AlAs heterosystem (GaAs: $\gamma=7.1$, $\gamma_{1}~=~2.554$; AlAs: $\gamma=3.76$, $\gamma_{1}=1.212$; $U_{0}=1.071$~eV~--- band offset). We have found the energy of the ground and excited states for the cubic, ellipsoidal, cylindrical, and pyramidal GaAs QD as a function of its volume. We note that the volume of a spherical QD, $\frac{4}{3} \pi R^{3} $, is equal to the volume of the other QDs.

In figures~\ref{fig1}--\ref{fig4}, dashed lines indicate the energy of the first three hole levels with quantum numbers $f=3/2$, $L=0$, $n=1$; $f=3/2$, $L=1$, $n=1$; $f=5/2$, $L=1$, $n=1$. The hole energy spectrum transforms in the cubic QD (figure~\ref{fig1}). In this case, the energies of the first two levels (solid lines~1, 2) are larger than the energies of the similar levels in the spherical QD. Solid lines 3 and 4 correspond to two- and fourfold degenerate levels obtained by splitting the second excited hole energy level of the spherical QD. Further, we consider the states with $n=1$ only, so this quantum number is omitted in the expressions of wave functions.

In figure~\ref{fig2} solid lines 1--3 correspond to the energy of the three first levels of holes in the ellipsoidal QD. As seen from the figure, changing the shape from a spherical to an ellipsoidal one is not followed by splitting any levels.

In a cylindrical QD (figure~\ref{fig3}), characterized by axial symmetry, three levels of the spherical QD (dashed lines) correspond to six energy levels (solid lines). In figure~\ref{fig3}, lines 1 and 2 indicate the energies of double degenerate levels corresponding to the ground state of the spherical QD, 3 and 4 denote the energies of double degenerate excited levels, lines 5 and 6 indicate the energies, respectively, of four- and twofold degenerate ones.

The calculations show that the hole energy spectrum in a tetrahedron-shaped QD differs from the hole spectrum of a spherical QD (figure~\ref{fig4}) more than in the case of the above mentioned shapes. Like in the previous figures, dashed lines represent the first three energy states of a spherical QD and solid lines denote the energy states obtained in a tetrahedral QD after splitting.

It is seen from the figures that for a QD of any form, the energies of all levels decrease with increasing the QD volume. Also, we see that for all shapes of QDs in the considered size limit, the corresponding energy levels are higher than those in a spherical QD. This result can be explained by stronger confinement of charges in the QDs of interest than in the spherical QD.

The wave functions of the hole states are much different even qualitatively in these types of QDs. For example, in a cubic QD at $V=34$~nm$^3$, the wave functions of the states with different energies are written as
%
\begin{align*}
E_{1} \; -\; \Psi _{1} & = v _{\frac{3}{2} ,0,\frac{3}{2} } \,,  \\
\quad \quad \Psi _{2}  & = v _{\frac{3}{2} ,0,-\frac{3}{2} } \,, \\
\quad \quad \Psi _{3}  & = v _{\frac{3}{2} ,0,-\frac{1}{2} } \, , \\
\quad \quad \Psi _{4}  & = v _{\frac{3}{2} ,0,\frac{1}{2} } \, , \\
%
E_{2} \; -\; \Psi _{5} & = -0.99v _{\frac{3}{2} ,1,\frac{3}{2} } -0.15v _{\frac{5}{2} ,1,-\frac{5}{2} } -0.08v _{\frac{5}{2} ,1,\frac{3}{2} } \, , \\
\quad \quad \Psi _{6}  & = -0.99v _{\frac{3}{2} ,1,-\frac{3}{2} } +0.08v _{\frac{5}{2} ,1,-\frac{3}{2} } +0.15v _{\frac{5}{2} ,1,\frac{5}{2} } \, , \\
\quad \quad \Psi _{7}  & = 0.98v _{\frac{3}{2} ,1,-\frac{1}{2} } -0.11v _{\frac{3}{2} ,1,\frac{1}{2} } +0.14v _{\frac{5}{2} ,1,-\frac{1}{2} } +0.02v _{\frac{5}{2} ,1,\frac{1}{2} } \, , \\
\quad \quad \Psi _{8}  & = 0.11v _{\frac{3}{2} ,1,-\frac{1}{2} } +0.98v _{\frac{3}{2} ,1,\frac{1}{2} } +0.02v _{\frac{5}{2} ,1,-\frac{1}{2} } -0.14v _{\frac{5}{2} ,1,\frac{1}{2} } \, ,\\
%
E_{3} \; -\; \Psi _{9}  & = 0.07v _{\frac{3}{2} ,1,-\frac{3}{2} } +0.99v _{\frac{5}{2} ,1,-\frac{3}{2} } -0.05v _{\frac{5}{2} ,1,\frac{5}{2} } \, ,  \\
\quad \quad \Psi _{10}  & = -0.07v _{\frac{3}{2} ,1,\frac{3}{2} } -0.05v _{\frac{5}{2} ,1,-\frac{5}{2} } +0.99v _{\frac{5}{2} ,1,\frac{3}{2} } \, , \\
%
E_{4} \; -\; \Psi _{11} & = 0.15v _{\frac{3}{2} ,1,-\frac{3}{2} } +0.04v _{\frac{5}{2} ,1,-\frac{3}{2} } +0.99v _{\frac{5}{2} ,1,\frac{5}{2} } \, , \\
\quad \quad \Psi _{12}  & = 0.15v _{\frac{3}{2} ,1,\frac{3}{2} } -0.99v _{\frac{5}{2} ,1,-\frac{5}{2} } -0.04v _{\frac{5}{2} ,1,\frac{3}{2} } \, ,  \\
\quad \quad \Psi _{13}  & = -0.1v _{\frac{3}{2} ,1,-\frac{1}{2} } +0.1v _{\frac{3}{2} ,1,\frac{1}{2} } +0.7v _{\frac{5}{2} ,1,-\frac{1}{2} } +0.7v _{\frac{5}{2} ,1,\frac{1}{2} } \, ,  \\
\quad \quad \Psi _{14}  & = -0.1v _{\frac{3}{2} ,1,-\frac{1}{2} } -0.1v _{\frac{3}{2} ,1,\frac{1}{2} } +0.7v _{\frac{5}{2} ,1,-\frac{1}{2} } -0.7v_{\frac{5}{2} ,1,\frac{1}{2} } \, .
\end{align*}

The calculations make it possible to obtain the wave functions of the states for all the considered shapes and volumes of QDs.

\section{The interlevel absorption coefficient}

We consider the case when linearly polarized light falls on the QD along the z-axis which coincides with the axis of symmetry of the QD. Then, the dipole momentum of transition between the $i\alpha $ and $j\beta $ states can be presented as
\begin{equation} \label{EQ__11_} M_{i\alpha ,j\beta } =\left\langle j\beta \left|ez\right|i\alpha \right\rangle =\int \Psi _{j\beta }^{*} (x,y,z) \ ez \ \Psi _{i\alpha } (x,y,z)\rd V ,
\end{equation}
\noindent and the transition oscillator strength is written as

\begin{equation} \label{EQ__12_} f_{i\alpha ,j\beta } =\frac{2m_{0} }{e^{2} \hbar ^{2} } (E_{j} -E_{i} )\left|M_{ia,j\beta } \right|^{2} . \end{equation}

The light absorption coefficient caused by the hole interlevel transition from the $E_{i} $ level into another energy level $E_{j} $ is defined as
\begin{equation} \label{EQ__13_} \alpha \left(\omega \right)=\omega \sqrt{\frac{\mu _{0} }{\varepsilon _{0} \varepsilon } } \sum _{i,j}\sum _{\alpha ,\beta }\frac{\sigma \left|M_{i\alpha ,j\beta } \right|^{2} \hbar \Gamma _{i\alpha ,j\beta } }{(E_{j} -E_{i} -\hbar \omega )^{2} +\left(\hbar \Gamma _{i\alpha ,j\beta } \right)^{2} } \,, \end{equation}
\noindent where $\alpha =1,2, \ldots, s_{\alpha } $, $\beta =1,2, \ldots, s_{\beta }$, $s_{\alpha }$ is the multiplicity of the degenerate level $i$, $s_{\beta }$ is the multiplicity of the degenerate level $j$, $\omega $ is the frequency of electromagnetic wave, $\varepsilon _{0} $ is the vacuum permittivity, $\mu _{0} $ is the vacuum permeability, $\varepsilon$ is the relative permittivity of a QD, $\hbar \Gamma _{i\alpha ,j\beta } $ is the relaxation energy due to the electron-phonon interaction and other scattering factors. The QD charge density $\sigma $ is chosen on the assumption that there can be only one electron in a QD, so $\sigma =3/\left(4\pi a^{3} \right)$.

According to the above formulas, we calculate the probability of optical transitions, the transition oscillator strength, and the light absorption coefficient for different geometric shapes of a QD. These physical quantities are compared with the corresponding parameters of the spherical-symmetry QD, for which transitions are allowed only between the levels under the following conditions:
\begin{itemize}
  \item transitions between states of equal parity are possible if $f'-f=\pm 1$, $M'-M=0,\pm 1$;
  \item transitions between odd and even states are possible if $f'-f=0$, $M'-M=0,\pm 1$.
\end{itemize}

\begin{figure}[!b]
\centerline{
\includegraphics[width=0.65\textwidth]{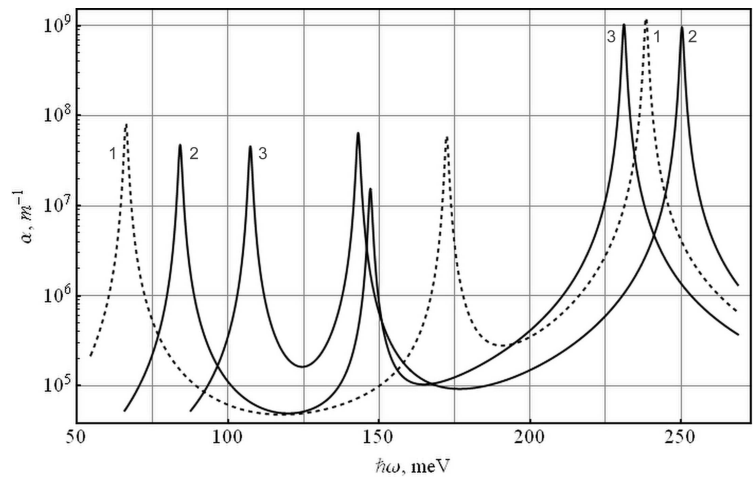}
}
	\caption{Absorption coefficients of interlevel optical transitions in a cubic (2) and a pyramidal (3) QDs in logarithmic scale. Dashed lines (1) correspond to a spherical QD.}
	\label{fig5}
\end{figure}

Equation~\eqref{EQ__13_} allows one to determine the dependence of the absorption coefficient on the frequency of external electromagnetic wave for transitions of the charge between the lowest levels for all the QD shapes. In figure~\ref{fig5} we present the dependence $\alpha =\alpha (\omega )$ at $V=34nm^3 $. Dashed lines correspond to the function $\alpha (\omega )$ for a spherical QD.

It is obvious that for spherical [figure~\ref{fig5}, (1)] and cubic [figure~\ref{fig5}, (2)] QD shapes, not only the absorption maxima differ, but also their energy, which is due to the difference between electron transition energies at a given size of the QD. So, if the absorption maximum for the two lowest levels in a spherical QD is equal to $8\cdot 10^{7}$~m$^{-1} $ with the energy 65~meV, then the cubic QD value of the corresponding interlevel absorption coefficient is $4.8\cdot 10^{7}$~m$^{-1} $ with the energy 85~meV, i.e., in a cubic QD the absorption maximum shifts towards higher energies with respect to a spherical QD. The absolute value of the maximum absorption coefficient between the second and the third lowest levels is lower (respectively, $5.6\cdot 10^{7}$~m$^{-1} $ and $1.5\cdot 10^{7}$~m$^{-1} $ for the sphere and cube). Besides, in the cubic QD, one observes the absorption peak shift towards lower energies. As to the third absorption band, which is caused by hole transitions between the first and third hole lowest levels, the maximum $\alpha (\omega )$ is the largest for all QD shapes (about $1.2\cdot 10^{9}$~m$^{-1} $).

\begin{figure}[!t]
\centerline{
\includegraphics[width=0.65\textwidth]{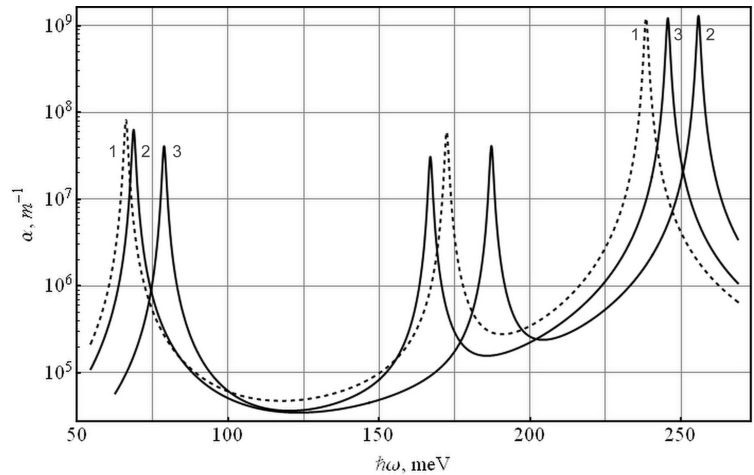}
}
	\caption{Absorption coefficients of interlevel optical transitions in an ellipsoidal (2) and a cylindrical (3) QDs in logarithmic scale.}
	\label{fig6}
\end{figure}

Comparing the absorption rate for interlevel optical transitions in QDs of ellipsoidal and spherical shapes [figure~\ref{fig6}, (2)], we can see that the maxima of the absorption coefficient are almost the same ($6.4\cdot 10^{7}$~$m^{-1} $, $8\cdot 10^{7}$~$m^{-1} $) for a transition of the charge between the two first levels, and the energies almost coincide (respectively, 70 and 65~meV). A small difference between these physical parameters of QDs can be explained by similarity of shapes between the sphere and the ellipsoid. It should be noted that all the maxima of the interlevel absorption coefficient in the ellipsoidal QD are shifted towards higher energies in comparison with the spherical QD.

As concerns the QD of cylindrical shape [figure~\ref{fig6}, (3)], the maxima of the light absorption coefficient in the hole transitions between the first and the second, or between the second and the third lowest energy levels are close to the respective maxima of the spherical QD ($4\cdot 10^{7}$~m$^{-1} $ and $3.1\cdot 10^{7}$~m$^{-1} $). The energies at which absorption occurs in the cylindrical QD (80 and 165~meV) are also close to the corresponding energies of the spherical QD (65 and 170~meV).

The QD of a regular tetrahedron shape differs greatly from the shape of the sphere. Therefore, the absorption coefficient of hole interlevel transitions in these QDs is also different [Fig.~\ref{fig5}, (2)]. The maxima $\alpha (\omega )$ for the charge transition between the three lowest levels are different from the maxima of the spherical QD (105 and 145~meV for a tetrahedral QD, 65 and 170~meV for a spherical QD). The absorption peaks are shifted towards larger and smaller energies, respectively, in comparison with the absorption peaks of a spherical QD.

The performed calculations based on the measurements $\alpha =\alpha (\omega )$, allow one to establish the shape of a QD which, as one can see, has a significant impact on the optical properties of a heterostructure.

\section{Conclusions}

The effect of the shape on the hole energy spectrum of a GaAs/AlAs heterostructure was investigated in this work. One takes into account the valence band degeneracy of two adjacent materials at a $\Gamma$ point of the Brillouin zone. The problem is solved for spherical, cubic, ellipsoidal (spheroidal), cylindrical, and pyramidal QDs provided their volumes are the same.

There are two types of QDs in our investigation. Some of them (spherical, cubic, pyramidal QDs) are characterized by one size parameter (radius or length of the edge). This parameter determines the volume of the QD. The volume of the other QDs is determined by means of two characteristic linear dimensions. Particularly, a QD of spheroidal shape is characterized by the length of small ($a$) and major ($b$) semiaxes, and a cylindrical QD is characterized by height ($h$) and diameter ($d$). The analysis shows that the potential $W$~\eqref{EQ__6_} can be considered as small perturbation for $15~\text{nm}^{3} \leqslant V\leqslant 530~\text{nm}^{3}$ if ratios $b/a\leqslant 1.5$; $h/d\leqslant 1.25$. These ratios of a spheroidal and cylindrical QDs remain constant and do not depend on size \-changes.

We can choose $\lambda  = \left\langle {1s\left| {U(\vec r) - {U_s}(r)} \right|1s} \right\rangle / U_0$ as a small parameter. When we change the volume of a QD in the range $15~\text{nm}^{3} \leqslant V\leqslant 530~\text{nm}^{3}$, the parameter $\lambda$ changes as: $0.0221122 \leqslant \lambda \leqslant 0.0676337$ (cubic shape), $0.00652098 \leqslant \lambda \leqslant 0.0181243$ (spheroidal shape), $0.0164079 \leqslant \lambda \leqslant 0.0408119$ (cylindrical shape), and $0.0291535 \leqslant \lambda \leqslant 0.0571925$ (pyramidal shape).

The precision of calculations is a very significant issue in any problem where one uses the methods of perturbation theory. In our approach, the improvement of calculation precision requires a great number of terms in the sum~\eqref{EQ__7_}. Computation time restricts the matrix order. Thus, one should choose a matrix with the minimum required number of matrix elements to find a real hole energy spectrum. Our calculations show that the energy of the ground and some lower excited levels will be fairly correct when we consider a $14\times14$ matrix. For instance, for a pyramidal QD with $V=34$~nm$^{3} $ (it corresponds to the $R=2$~nm spherical QD) the diagonalization of the $18\times18$ matrix makes it possible to obtain more exact hole energy levels (the data of diagonalization of the $14\times14$ matrix are presented in brackets, energy values are given in~meV):
\begin{gather*}
E_{1} =291\, (293),\\
E_{2} =318\, (320),\\
E_{3} =399\, (404),\\
E_{4} =429\, (435),\\
E_{5} =553\, (553),\\
E_{6} =572\, (575),\\
E_{7} =635\, (631).
\end{gather*}
\noindent One can see that the relative error of the calculations is less than~$1.3\% $.

The energy spacing between the hole adjacent levels is sufficiently large at such a QD volume of any shape. In spite of a simultaneous existence of QDs with close but still different volumes under real physical conditions, one can distinguish the examined hole levels in the analysis of a frequency dependence of the interlevel hole absorption coefficient. We can draw a conclusion that the values of maxima and the energies of the peaks of the hole interlevel absorption coefficient depend on the QD shape.

The obtained data can be used to determine the shape of a GaAs QD in the AlAs matrix after the comparison of the experimental and theoretical data.


\ukrainianpart

\title{Вплив форми квантової точки гетеросистеми GaAs/AlAs на міжрівневе діркове поглинання світла}
\author{В.І.~Бойчук, І.В.~Білинський, О.А.~Сокольник, І.О.~Шаклеіна}
\address{Дрогобицький державний педагогічний університет ім. Івана Франка, Кафедра теоретичної фізики,  \\ вул. Стрийська, 3,  82100 Дрогобич, Україна}

\makeukrtitle

\begin{abstract}
\tolerance=3000%
Для квантової точки GaAs у напівпровідниковій матриці AlAs досліджено вплив форми квантової точки на енергетичний спектр дірки та оптичні властивості, зумовлені міжрівневими переходами заряду, на основі гамільтоніана $4\times4$. Обчислення проводились за допомогою теорії збурень з врахуванням гібридизації станів для кубічної, еліпсоїдальної, циліндричної та тетраедричної форми при зміні об'єму квантової точки. На основі проведених обчислень енергії та визначених хвильових функцій станів дірки встановлено правила відбору та досліджено залежність коефіцієнту міжрівневого діркового поглинання світла від форми та об'єму квантової точки.

\keywords багатозонне наближення, оптичні переходи, сили осцилятора, коефіцієнт поглинання

\end{abstract}


\end{document}